\documentclass[10pt,aps,prx,twocolumn,superscriptaddress,nofootinbib]{revtex4-2}
\usepackage[english]{babel}
\usepackage[utf8]{inputenc}
\usepackage[T1]{fontenc}
\usepackage{graphicx}
\usepackage{amsmath,amssymb,amsfonts}
\usepackage{xcolor}
\usepackage{physics}
\usepackage{subcaption}
\usepackage{ulem}
\usepackage{ragged2e}
\usepackage{cmbright}
\usepackage{comment}
\usepackage[colorlinks=true,linkcolor=blue,citecolor=blue,urlcolor=blue]{hyperref}

\begin{document}

\title{Time-bin qubit architecture using quantum Hall edge channels}

\author{David Pomaranski}
\email{david@ap.t.u-tokyo.ac.jp}
\affiliation{The University of Tokyo, Tokyo, Japan}
\affiliation{RIKEN Center for Emergent Matter Science (CEMS), Wako, Saitama, Japan}

\author{Michihisa Yamamoto}
\email{michihisa.yamamoto@riken.jp}
\affiliation{The University of Tokyo, Tokyo, Japan}
\affiliation{RIKEN Center for Emergent Matter Science (CEMS), Wako, Saitama, Japan}

\date{\today}

\begin{abstract}
We present the basic elements for a modular architecture for time-bin encoded qubits based on quantum Hall edge channels, forming the foundation of a scalable electronic quantum information platform named TEMPO (Time-binned Electronic Modular Platform for Qubits). Quantum states are encoded in temporally separated edge magnetoplasmon (EMP) wave packets propagating along a single chiral edge, eliminating the need for spatial path separation and enhancing coherence. The platform supports full qubit operations—including initialization, phase modulation, readout, and two-qubit entangling gates—by leveraging dynamically tunable quantum point contacts and electrostatic control of interferometric loops. We consider the linear dispersion and gate-induced velocity control on EMP propagation and describe strategies for maintaining waveform integrity. Various single-electron sources, including ohmic injection and capacitive excitation, are discussed in the context of coherence. Multi-qubit operations are enabled through synchronized injection and engineered Coulomb interactions between adjacent channels, while single-qubit readout is addressed via spin-based or capacitive charge sensors. Our approach integrates gate-tunable coherent control of chiral edge states, offering a comprehensive pathway toward scalable electron quantum optics in solid-state platforms.
\end{abstract}

\maketitle

\section{Introduction}

Quantum information processing requires a platform that can coherently manipulate quantum states while minimizing the impact of environmental noise and material imperfections. A major challenge for many qubit implementations, including superconducting, spin-based, and semiconductor qubits, lies in mitigating decoherence, while integrating a large number of qubits. Flying qubit architectures, which encode quantum information in propagating modes, have been proposed as a promising alternative. This approach offers flexibility in routing, coupling, and scaling~\cite{ionicioiu_quantum_2001, yamamoto_electrical_2012, pomaranski_semiconductor_2024}.

Quantum Hall edge channels provide a compelling platform for flying qubits due to three key properties: (i) topological protection suppresses disorder-induced backscattering, (ii) the ability to host edge magnetoplasmons (EMPs) and collective quasiparticle excitations, which extend the phase coherence beyond that of single-electron excitations~\cite{hiyama_edge-magnetoplasmon_2015}, and (iii) precise phase control of GHz electron dynamics by electrostatic gating~\cite{ji_electronic_2003}. Single-electron sources~\cite{feve_-demand_2007} and electron interferometers~\cite{neder_interference_2007} have demonstrated coherent control over individual electronic wave packets~\cite{ito_coherent_2021,ouacel_electronic_2025,assouline_emission_2023}. 
Despite these advances, early flying qubit proposals based on two-path interferometers faced serious challenges with scalability. In two-dimensional electron systems (2DES), the device topology often required coherent electrons to converge at a central node—typically a small metallic island used for readout. While this allowed for interference-based measurements within the device, the small size and metallic nature of the node disrupted phase coherence. As a result, these architectures could not be naturally extended into larger circuits or used for sequential quantum operations~\cite{bordone_quantum_2019}.

Although decoherence mechanisms have not been systematically characterized for time-resolved electron wave packets, it's known that electron wave packets, such as those based on EMPs, can exhibit significantly longer coherence times due to the suppression of energy relaxation. EMPs on quantum Hall edge channels may serve as ideal flying qubits if scalable circuits are realized in these systems. Compared to bare electron excitations EMPs are collective charge-density modes whose dynamics are governed by long-range Coulomb interactions. This collective nature allows EMPs to average over local potential fluctuations, making them less susceptible to short-range disorder and gate-induced noise~\cite{hiyama_edge-magnetoplasmon_2015, hashisaka_distributed-element_2013}. As a result, EMPs can propagate over tens to hundreds of microns with minimal energy relaxation, particularly at low filling factors such as $\nu = 1$~\cite{kamata_fractionalized_2014}. However, EMPs are still subject to dephasing arising from fluctuations in the electrostatic potential along the edge channel. This dephasing may limit the coherence of phase-sensitive interference. Our novel time-bin encoding scheme helps mitigate this issue by storing quantum information in the relative delay and phase between two well-separated temporal modes. Since both components of the time-bin qubit travel along the same physical path, they experience nearly identical environmental noise, leading to effective suppression of phase decoherence. This robustness to low-frequency potential fluctuations makes time-bin qubits a promising approach for preserving coherence in electron quantum optics.

Our proposed architecture, named TEMPO (Time-binned Electronic Modular Platform for Qubits) addresses many issues faced by previous proposals for quantum information processing utilizing quantum Hall edge channels by routing both basis states along the same physical edge channel with only a temporal offset. This modular design has the potential to serve as a foundation for scalable quantum logic using electron optics in solid-state platforms.

\section{Background and Motivation}

Flying electronic qubits encode quantum information in propagating degrees of freedom and offer a natural pathway to modular, on-chip quantum processing.~\cite{ionicioiu_quantum_2001, yamamoto_electrical_2012, pomaranski_semiconductor_2024} Early investigations in the quantum Hall regime implemented electronic analogs of optical interferometers using quantum point contacts (QPCs) as beam splitters~\cite{ji_electronic_2003, neder_interference_2007, roulleau_direct_2008, litvin_decoherence_2007}. These experiments demonstrated phase-coherent electron transport, Hong-Ou-Mandel (HOM) interference~\cite{bocquillon_coherence_2013}, and two-electron entanglement schemes~\cite{freulon_hong-ou-mandel_2015, marguerite_two-particle_2017}

The development of on-demand single-electron sources marked a major milestone for electron quantum optics. Key approaches include mesoscopic capacitors~\cite{feve_-demand_2007,mahe_current_2010}, Leviton sources via Lorentzian voltage pulses~\cite{levitov_electron_1996, glattli_levitons_2017, dubois_integer_2013, jullien_quantum_2014}, which enable deterministic injection of single-electron wave packets into chiral edge states with tunable energy, timing, and pulse width. Surface acoustic wave (SAW) techniques have also enabled the transport of single electrons along depleted channels~\cite{ito_coherent_2021,wang_generation_2022}, demonstrating single-electron transfer between distant quantum dots on macroscopic length scales.~\cite{hermelin_electrons_2011, mcneil_-demand_2011}

Time-bin encoding was originally developed in photonic systems as a robust method for quantum communication in dispersive and noisy environments. These photonic qubit architectures use unbalanced interferometers to create early and late temporal modes that form the logical basis states, making them less sensitive to differences in phase fluctuations arising in spatially separated paths and allowing for long-distance coherent transmission along an optical fiber~\cite{brendel_pulsed_1999, marcikic_distribution_2004}. 

Recent efforts in quantum Hall systems have turned toward exploring edge magnetoplasmons (EMPs)—collective charge-density excitations that propagate chirally at GHz frequencies—as carriers of quantum information for electronic flying qubit architectures. These EMPs arise naturally in the presence of strong magnetic fields and exhibit properties that are well-suited for time-bin encoding, including high temporal coherence, robustness against local disorder, and compatibility with fast, gate-based modulation~\cite{kamata_voltage-controlled_2010, hashisaka_distributed-element_2013}. Conceptual proposals~\cite{ionicioiu_quantum_2001, haack_coherence_2011} have laid the groundwork for such schemes, motivating the development of qubit architectures that leverage the temporal degree of freedom of EMPs for initialization, manipulation, and readout. These characteristics make EMP-based time-bin encoding a promising approach for scalable quantum information processing in solid-state platforms.

\section{Architecture Overview}

\begin{figure}[ht]
    \centering
    \begin{subfigure}[b]{0.99\linewidth}
        \centering
        \includegraphics[width=\linewidth]{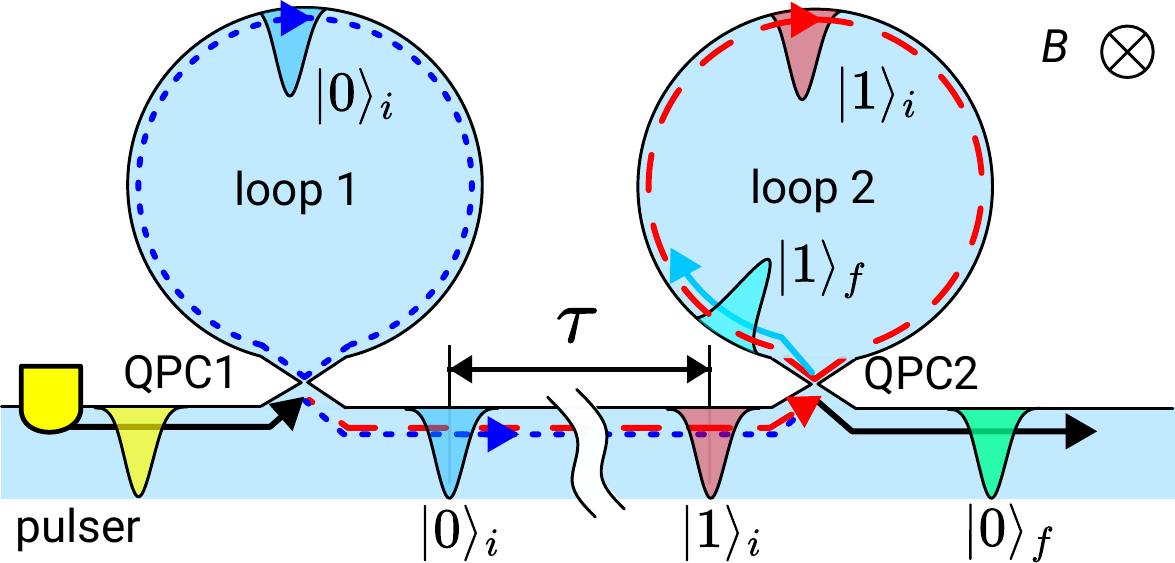}
        \caption{Device-level architecture of the time-bin qubit using chiral edge channels in the quantum Hall regime. EMP wave packets are injected via a pulser and propagate unidirectionally along the edge of a 2DEG. Quantum point contacts (QPC1 and QPC2) define two loops, enabling state preparation and phase modulation. The magnetic fluxes \(\varphi_1\) and \(\varphi_2\) through each loop control the relative phase between different paths and are tunable via electrostatic side gates that locally deplete the 2DEG, thus modifying the effective area enclosed by the edge channel.}
        \label{fig:arch_a}
    \end{subfigure}
    \hfill
    \begin{subfigure}[b]{0.99\linewidth}
        \centering
        \includegraphics[width=\linewidth]{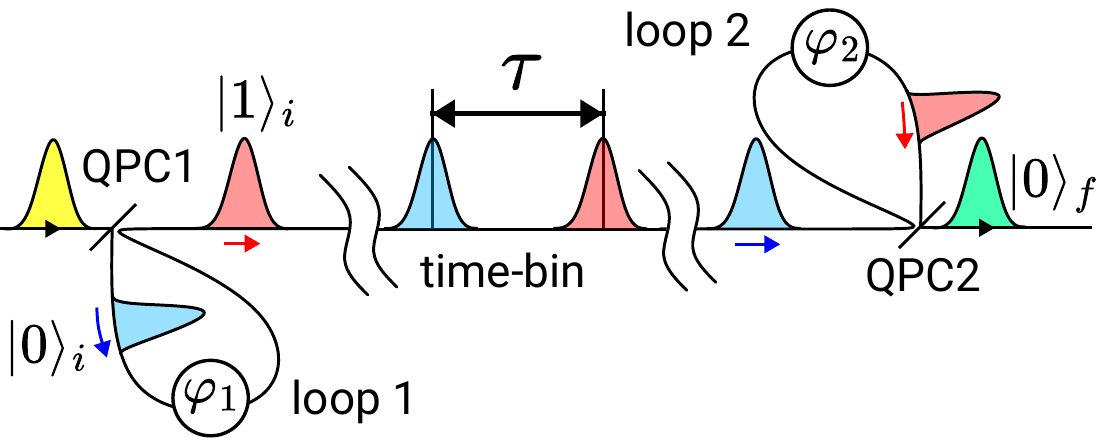}
        \caption{Time-bin qubit operation illustrated in a Mach-Zehnder–like schematic. QPC1 and QPC2 act as dynamically controlled beam splitters, synchronizing the splitting and recombination of wave packets. Each QPC can be set to 50:50 transmission (or another ratio depending on the quantum operation). Varying the enclosed flux \(\varphi_1\) or \(\varphi_2\) enables phase tuning between the two arms, allowing standard Mach-Zehnder interference to be observed at the output.}
        \label{fig:arch_b}
    \end{subfigure}
    \caption{(a) Physical implementation of the time-bin flying qubit architecture. (b) Logical interferometric representation of the qubit operation in the time domain.}
    \label{fig:architecture}
\end{figure}

The device architecture is illustrated in Fig.~\ref{fig:architecture}. Panel~(a) shows a top-down schematic of a two-dimensional electron gas (2DEG), where the blue-shaded areas represent conducting regions defined by either mesa etching or electrostatic depletion via surface gates. When subjected to a strong perpendicular magnetic field (pointing into the page), electrons in the 2DEG form one-dimensional chiral edge channels that support unidirectional propagation along the device perimeter.

A localized excitation, labeled as the “pulser,” injects high-frequency voltage pulses into the 2DEG, exciting edge magnetoplasmon (EMP) wave packets that propagate chirally toward quantum point contact 1 (QPC1). QPC1 acts as a tunable beam splitter: when set to a transmission probability \( T = 0.5 \), it coherently splits the incoming wave packet equally into two partial waves. One wave is transmitted along the lower path directly toward QPC2, defining the logical basis state \(\lvert 0 \rangle\); the other is reflected into the upper interferometric loop (loop~1), defining the delayed state \(\lvert 1 \rangle\). These two components form a superposition of a two-path qubit state. After completing a single round trip in loop~1, the \(\lvert 1 \rangle\) component returns to QPC1. Moments prior to this event, QPC1 is dynamically opened to full transmission, allowing the \(\lvert 1 \rangle\) component to exit along the same edge channel as the earlier \(\lvert 0 \rangle\) component, but offset in time by \(\tau\). This synchronized emission defines a time-bin qubit, in which both states propagate along the same edge channel and are separated by a precisely defined time delay. To preserve their orthogonality in the time domain, the temporal width of the initial pulse must be smaller than the round-trip delay time \(\tau = L_1 / v_{\text{EMP}}\), where \(L_1\) is the perimeter of loop~1 and \(v_{\text{EMP}}\) is the velocity of EMP propagation.

To enable quantum control of the time-bin state and to perform phase-sensitive interference measurements, the architecture includes a second loop (loop~2) connected via QPC2. Initially, QPC2 is fully open so that the \(\lvert 0 \rangle\) component enters the loop. After completing a single round trip in loop~2, the delayed \(\lvert 1 \rangle\) component arrives at QPC2 simultaneously with the \(\lvert 0 \rangle\) component, but from opposite sides. At this moment, QPC2 is set to the beam-splitter condition \(T = 0.5\), enabling the two components to interfere at the output in a "time-domain analog" of a Mach–Zehnder interferometer, as illustrated in Fig.~\ref{fig:architecture}b. The phase difference between the two paths—controlled via local side gates that tune the magnetic fluxes \(\varphi_1\) and \(\varphi_2\) enclosed by loops~1 and 2—determines the interference pattern observed at the output, allowing for qubit state projection in an arbitrary basis.

This architecture also supports full single time-bin qubit operation by tuning both $T$ at the QPCs and the phase shift: initialization via EMP injection, coherent control through dynamic QPC gating and flux-tunable phase shifts, and projective readout or non-destructive measurement protocols.\cite{glattli_design_2020,thiney_-flight_2022} Its modular design allows for integration into more complex electron quantum optics circuits, making it a promising platform for scalable quantum information processing~\cite{pomaranski_semiconductor_2024}.

\subsection{Dispersion and decoherence}
To suppress pulse distortion and preserve the shape of single-electron wave packets, it is desirable for edge magnetoplasmons (EMPs) to propagate with a linear dispersion relation. In general, the dispersion of EMPs in a bare two-dimensional electron gas (2DEG) is weakly nonlinear due to long-range Coulomb interactions, with the form $\omega(k) \propto k \log(1/kd)$, where $k$ is the wave vector and $d$ is the distance to a nearby screening gate~\cite{fetter_edge_1985, volkov_edge_1988, aleiner_novel_1994}. This nonlinearity causes different frequency components of the wave packet to propagate at different velocities, leading to temporal broadening.

However, the precise form of the dispersion depends on the electrostatic environment. When a metallic gate is placed near the edge channel ($d \ll w$), where $w$ is the transverse spatial extent of the electric field (typically on the micron scale), the long-range Coulomb interaction is screened. In this regime, the dispersion becomes linear and takes the form $\omega(k) = \left( \sigma_{xy} / \epsilon \right) (d/w) \, k$~\cite{kamata_voltage-controlled_2010, hashisaka_distributed-element_2013}. Here, $\sigma_{xy}$ is the Hall conductance and $\epsilon$ is the dielectric constant of the surrounding material. This linear dispersion is advantageous for maintaining the shape and coherence of propagating wave packets.

In contrast, for $d \gg w$, screening is weak and the dispersion reverts to its nonlinear form. Thus, while top gates are essential for electrostatic manipulation and phase control, they must be carefully engineered to ensure global and uniform screening—ideally maintaining $d \ll w$—to preserve linear dispersion and minimize waveform distortion. These considerations highlight the critical role of gate geometry and material design~\cite{kumada_suppression_2020} in high-fidelity EMP-based quantum information architectures.

\subsection{Electron sources}

Edge magnetoplasmons (EMPs) can be excited either capacitively or via ohmic injection, enabling different forms of single-electron excitation. In the capacitive case, a fast voltage signal applied to a Schottky gate near the edge channel produces a transient displacement current \( I(t) = C \, dV(t)/dt \). Since the total integrated current is zero, this method injects charge-neutral excitations, corresponding to a pair of oppositely charged density fluctuations propagating along the edge. These wave packets, sometimes viewed as dipole-like plasmon modes, do not transfer net charge into the 2DEG. Although they have been used in various interference experiments, their coherence properties remain less well understood due to their neutral nature, but may have desirable properties.

In contrast, ohmic injection provides a route to creating well-defined, charge-carrying single-electron wave packets. Applying a voltage pulse to an ohmic contact modulates the local electrochemical potential, injecting excess charge (electrons or holes) into the edge channel. A positive pulse generates an electron-like excitation (quasiparticle), while a negative pulse produces a hole-like excitation. The resulting wave packets follow the envelope of the voltage drive, making them natural candidates for encoding quantum information in the time domain.

A particularly important class of such excitations is the Leviton—a minimal excitation formed by applying a quantized Lorentzian voltage pulse of the form \( V(t) \propto 1 / \left[(t - t_0)^2 + w^2\right] \) to an ohmic contact~\cite{dubois_integer_2013}. Levitons carry an integer charge and no accompanying electron-hole pairs, making them clean single-particle states with well-defined coherence. This purity enables their use in electronic quantum optics, where they can interfere with other flying qubits in beam splitter geometries~\cite{glattli_levitons_2017}.

The fidelity and coherence of these single-electron excitations have been characterized using quantum state tomography techniques. Jullien et al.~\cite{levitov_electron_1996} performed pioneering work demonstrating Wigner tomography of Levitons by analyzing shot noise in an electron beam splitter interferometer. Later, Bisognin et al.~\cite{bisognin_quantum_2019} and Fletcher et al.~\cite{fletcher_continuous-variable_2019} extended this to continuous-variable quantum tomography of current-carrying wave packets, enabling reconstruction of the energy and temporal structure of single-electron excitations with high resolution. These developments highlight the versatility of time-domain single-electron sources, and offer building blocks for flying qubit architectures. Their performance depends crucially on minimizing dispersion and dephasing during propagation and on precise waveform engineering at the source.

\subsection{Phase modulation}

The time-bin qubit architecture inherently supports phase control through the Aharonov-Bohm (AB) effect, which arises from the magnetic flux $\Phi_B$ enclosed by each of the loops. As the $\lvert 0 \rangle$ state propagates around this closed path while the $\lvert 1 \rangle$ state bypasses it, a relative phase difference is accumulated. This phase shift is given by $\Delta \varphi = \frac{q \Phi_B}{\hbar}$ where $q$ is the charge of the excitation (typically $-e$ for an electron or effective quasiparticle). In practice, the enclosed magnetic flux $\Phi_B$ can be dynamically tuned by adjusting the effective area of the loop. This is achieved via an electrostatic gate positioned near the edge of the loop (not shown in Figure \ref{fig:architecture}), which modifies the local confinement potential and thereby shifts the position of the edge channel. Such control enables fine tuning of the AB phase without requiring changes to the external magnetic field. 

This gate-tunable phase shift allows coherent manipulation of the relative phase between $\lvert 0 \rangle$ and $\lvert 1 \rangle$ components of the time-bin qubit, effectively enabling a $\hat{Z}$-rotation in the qubit Hilbert space. When combined with amplitude modulation at the first beam splitter (QPC1), which sets the superposition coefficients, this phase control provides a pathway toward arbitrary single-qubit gate operations within the time-bin encoding framework.

\subsection{Multi-qubit operations}

\begin{figure}[ht]
    \centering
    \includegraphics[width=0.99\linewidth]{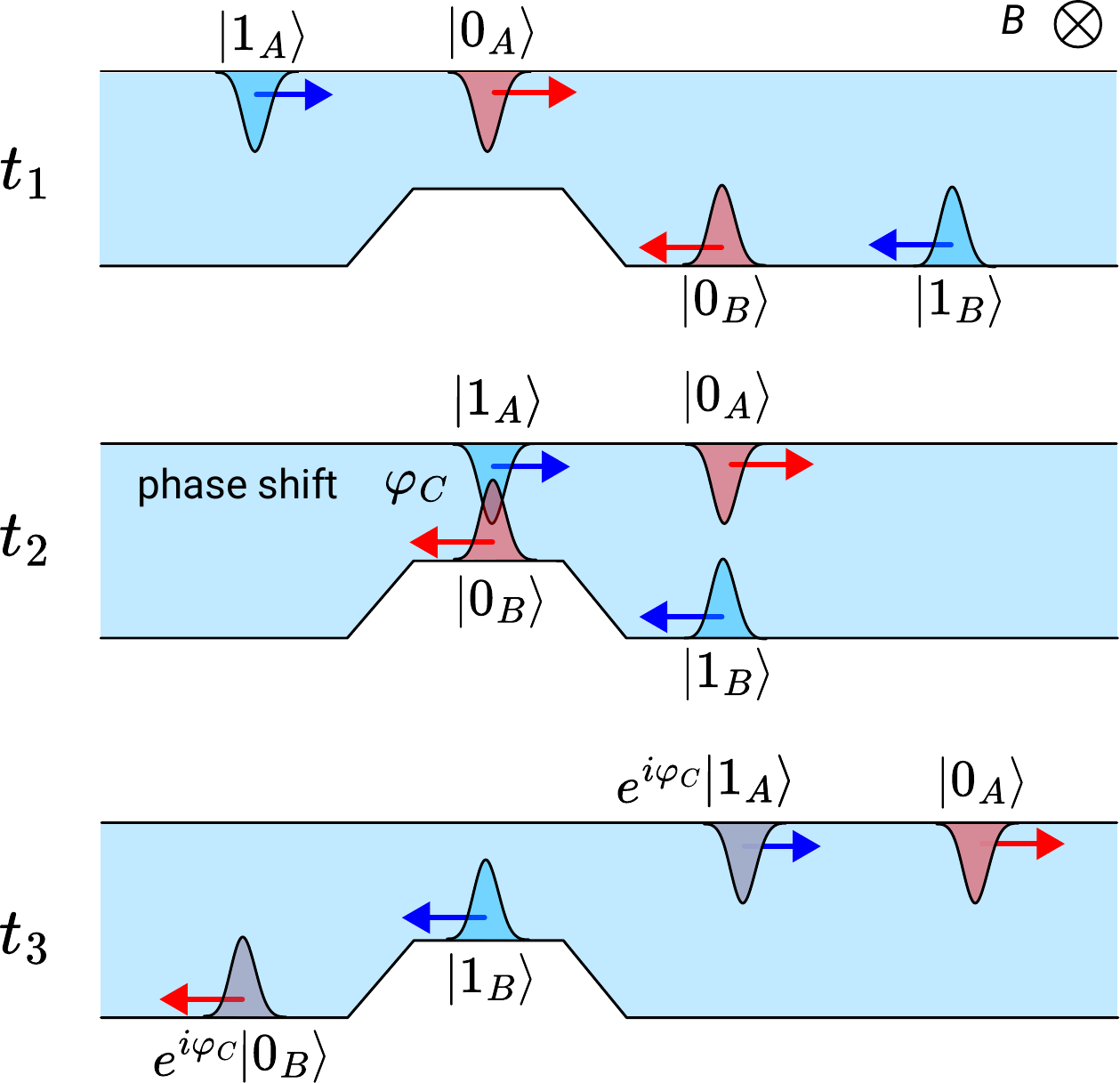}
    \caption{Schematic of a two-qubit interaction. Two time-bin qubits (A and B), encoded in counter-propagating edge channels, encounter an interaction region. Their respective wave packets \(|1_A\rangle\) and \(|0_B\rangle\) arrive simultaneously at time \(t_2\), leading to a controlled phase shift due to their mutual Coulomb interaction. The interaction strength—and resulting entangling phase—can be tuned by adjusting the spatial overlap, velocity, and confinement geometry of the interaction region.
    }
    \label{fig:fig2}
\end{figure}

A central requirement for scalable quantum information processing is the ability to perform entangling operations between flying qubits. In photonic systems, such operations are typically limited by weak nonlinearities and the absence of interactions. In contrast, electronic systems based on chiral edge channels naturally enable controlled interactions via the Coulomb force, offering an intrinsic mechanism for multi-qubit gates.

One strategy for realizing two-qubit gates involves capacitive Coulomb coupling between adjacent, counter-propagating edge channels. In the configuration shown in Fig.~\ref{fig:fig2}, two time-bin qubits are launched in opposite directions, such that the \(|1_A\rangle\) component of qubit A and the \(|0_B\rangle\) component of qubit B arrive simultaneously at a central constriction. Their overlap in this region induces a state-dependent phase shift \(\varphi_C\) due to their mutual interaction. By engineering the length, electrostatic potential, and timing of the interaction region, the resulting entangling operation can be tuned to implement a controlled-phase (CPHASE) gate.

As discussed by Glattli and Roulleau~\cite{glattli_levitons_2017}, such Coulomb-induced entangling operations are a promising path to implement flying-qubit versions of CNOT gates. By injecting synchronized Levitons into chiral edge channels, distant scattering between electrons—mediated purely by Coulomb interaction—can induce a conditional $\pi$-phase shift. This mechanism enables the creation of Bell states such as $\lvert 1_A \rangle \lvert 1_B \rangle + \lvert 0_A \rangle \lvert 0_B \rangle$, encoded in the joint time-bin basis of two flying qubits.

As highlighted in the review by Bäuerle et al.~\cite{bauerle_coherent_2018}, realizing two-qubit operations in ballistic electron systems remains a frontier challenge. Experimental progress in synchronized single-electron sources, precise control of edge state trajectories, and improved coherence times will be crucial for advancing beyond single-qubit manipulation toward scalable flying-qubit architectures.

\subsection{Readout}

An essential component of any quantum information architecture is the ability to perform reliable single-qubit readout. In the context of flying qubits encoded in EMPs, this corresponds to detecting the presence or absence of a single electron in a specific time bin with high fidelity. Several approaches have been proposed and experimentally pursued to achieve this goal with high fidelity.

One promising method leverages a nearby singlet-triplet (S–T) spin qubit implemented in a double quantum dot. When an EMP passes in close proximity to such a qubit, its transient electric field modifies the exchange interaction between the two dots. This leads to a measurable change in the spin qubit’s time evolution, effectively enabling it to function as a single-shot detector for EMPs~\cite{thiney_-flight_2022}.

In addition to spin-based detection, other approaches such as quantum point contact (QPC) charge sensors or mesoscopic capacitive probes have also been explored~\cite{bauerle_coherent_2018}. These techniques offer continuous or time-resolved detection of charge fluctuations and can, in principle, resolve individual electron wave packets with suitable bandwidth. Moreover, recent proposals suggest that non-demolition measurements of edge wave packets may be feasible using carefully designed readout architectures~\cite{glattli_design_2020}. In such schemes, the electron interacts with a detector multiple times while circulating in a loop, allowing information to be extracted without fully collapsing the wavefunction. This opens a path toward repeated measurements and real-time quantum feedback in flying qubit architectures.

Ultimately, high-fidelity readout of flying qubits remains a key technical challenge. The development of integrated, minimally invasive detectors that can operate on sub-nanosecond timescales will be critical for scaling up electron quantum optics platforms. Coupling robust single-shot detectors with time-bin architectures promises a path toward full qubit state tomography and real-time quantum feedback protocols in solid-state systems.

\section{Conclusion and outlook}

We have proposed a hardware platform for time-bin encoded qubits based on quantum Hall edge channels, combining concepts from photonic quantum communication with the intrinsic advantages of electronic systems. The architecture utilizes edge magnetoplasmons (EMPs) to encode quantum information in temporally separated wave packets that propagate along a single chiral edge. By using dynamically controlled quantum point contacts and electrostatic phase modulators, we demonstrate how time-bin qubits can be initialized, manipulated, and read out with high fidelity.

Our design offers several key advantages over conventional two-path flying qubit schemes: (i) the time-bin basis circumvents the need for separate interference paths, thereby reducing sensitivity to fabrication disorder and environmental noise; (ii) modular loops allow for reconfigurable phase control and synchronization; and (iii) the architecture naturally supports integration of multiple qubits via Coulomb interaction and synchronized single-electron injection.

We also described the role of dispersion and decoherence in EMP propagation, strategies for mitigating waveform distortion, and advances in detector development enabling full quantum state characterization. Together, these components form a comprehensive toolbox for coherent quantum control of single-electron wave packets.

Looking forward, scaling up this architecture will require advances in several areas. First, achieving universal quantum logic will necessitate high-fidelity two-qubit gates based on tunable Coulomb interactions. Second, integrated and minimally invasive detection schemes with single-electron sensitivity and sub-nanosecond resolution must be further developed. Time-bin architectures such as TEMPO represent a promising direction in the development of electronic quantum information processors. By leveraging the chiral, coherent nature of quantum Hall edge states in a temporally encoded framework, this platform offers a scalable path toward on-chip quantum optics with electrons.

\section{Acknowledgements}
We acknowledge Japan Society for the Promotion of Science, Grant-in-Aid for Scientific Research S (grant number JP24H00047), CREST-JST (grant number JPMJCR1675) and JST Moonshot (grant numbers JPMJMS226B-4). 

\bibliographystyle{apsrev4-2}
\bibliography{references}

\clearpage

\end{document}